\documentclass[letter,10pt]{article} 
\usepackage{color,amssymb,amsthm,amsmath,amsxtra,fullpage,graphicx,hyperref} 

\usepackage{abstract}

\def\bfS{{\textbf{S}}}
\def\mod{\text{ mod }}

\numberwithin{equation}{section}

\begin{document}

\def\bfS{{\textbf{S}}}

\def\todo#1{\textcolor{red}{\textbf{**** TODO -- #1 ****}}}

\renewcommand{\qed}{\nobreak \ifvmode \relax \else
      \ifdim\lastskip<1.5em \hskip-\lastskip
      \hskip1.5em plus0em minus0.5em \fi \nobreak
      \vrule height0.75em width0.5em depth0.25em\fi}

\newtheorem{theorem}{Theorem}[section]
\newtheorem{lemma}[theorem]{Lemma}
\newtheorem{conjecture}[theorem]{Conjecture}
\newtheorem{proposition}[theorem]{Proposition}
\newtheorem{corollary}[theorem]{Corollary}

\renewcommand{\qed}{\nobreak \ifvmode \relax \else
      \ifdim\lastskip<1.5em \hskip-\lastskip
      \hskip1.5em plus0em minus0.5em \fi \nobreak
      \vrule height0.75em width0.5em depth0.25em\fi}

\theoremstyle{definition}
\newtheorem{example}[theorem]{Example}
\newtheorem{definition}[theorem]{Definition}	
\newtheorem{construction}[theorem]{Construction}
%
%

\centerline{{\LARGE A Construction for Perfect Periodic Autocorrelation Sequences}}
\medskip
\centerline{\large Samuel T. Blake, Andrew Z. Tirkel}
\smallskip
\centerline{\large \it School of Mathematical Sciences, Monash University, Australia}
\bigskip

\begin{abstract}
	\noindent{\sc Abstract.}  We introduce a construction for perfect periodic autocorrelation sequences 
	over roots of unity. The sequences share similarities to the perfect periodic sequence constructions 
	of Liu, Frank, and Milewski.
\end{abstract}

Perfect periodic autocorrelation sequences see applications in many areas, including 
spread spectrum communications \cite{Simon1985}, channel estimation and 
fast start-up equalization \cite{Milewski1983}, pulse compression radars \cite{Farnett1990}, sonar 
systems \cite{Xu2011}, CDMA systems \cite{Ipatov2005}, 
system identiÞcation \cite{VanSchyndel2001}, and watermarking \cite{Tirkel1993}.\\

There exists a number of known constructions for perfect sequences over roots
of unity. These include Frank sequences of length $n^2$ over $n$ roots of unity
\cite{Heimiller1961}\cite{Frank1962}, Chu sequences of length $n$ over $n$ roots of unity for 
$n$ odd and length $n$ over $2n$ roots of unity for $n$ even \cite{Chu1972}, 
Milewski sequences of length $m^{2k+1}$ over $m^{k+1}$ roots of unity \cite{Milewski1983},
Liu-Fan sequences of length $n$ over $n$ roots of unity for $n$ even \cite{Liu2004}. Other
sequence constructions exist \cite{Ipatov1979}\cite{Alltop1980}\cite{Lewis1982}\cite{Kumar1985}
\cite{Gabidulin1993}\cite{Mow1993}. This construction presented in this paper is similar to 
the ZCZ sequence construction by the authors \cite{Blake2012}. \\

The periodic cross-correlation of the sequences, $\textbf{a} = \left[a_0,a_1,\cdots, a_{n-1}\right]$ and 
$\textbf{b} = \left[b_0, b_1,\cdots, b_{n-1}\right]$, for shift $\tau$ is given by 
$$\theta_{\textbf{a},\textbf{b}}(\tau) = \sum_{i=0}^{n-1}a_i b_{i+\tau}^*,$$ where $i+\tau$ is computed
modulo $n$. The periodic autocorrelation of a sequence,
\textbf{s} for shift $\tau$ is given by $\theta_{\textbf{s}}(\tau) = \theta_{\textbf{s},\textbf{s}}(\tau)$. For 
$\tau \neq 0 \mod n$, $\theta_{\textbf{s}}(\tau)$ is called an \textit{off-peak} autocorrelation. A sequence 
is {\it perfect} if all off-peak autocorrelation values are zero.\\

The periodic autocorrelation of a sequence, $\textbf{s} = [s_0, s_1, \cdots, s_{ld^2-1}]$, can be expressed 
in terms of the autocorrelation and cross-correlation of an array {\it associated} with \textbf{s} \cite{Heimiller1961}\cite{Frank1962}\cite{Mow1993}.  The sequence 
\textbf{s} has the {\it array orthogonality property} (AOP) for the {\it divisor} $d$, if the array 
\textbf{S} associated with \textbf{s} has the following two properties:
\begin{enumerate}
\item For all $\tau$, the periodic cross-correlation of any two distinct columns of \textbf{S} is zero.
\item For $\tau\neq0$, the sum of the periodic autocorrelation of all columns of \textbf{S} is zero. 
\end{enumerate}

Any sequence with the AOP is perfect \cite{Mow1993}.\\

In most perfect sequence constructions, one proves the sequence has perfect autocorrelation by reducing the 
autocorrelation to a Gaussian sum. A Gaussian sum is given by 
$\sum_{k=0}^{n-1}\omega^{q k}$, where $\omega = e^{2 \pi \sqrt{-1}/n}$ and $q \in \mathbb{Z}$. 
If $q \neq 0 \mod n$, then the sum is zero. \\

We present a construction for perfect sequences over roots of unity. 
Let \textbf{s} be a sequence of length $4mn^{k+1}$ over $2mn^k$ roots of 
unity, where $n,m,k \in \mathbb{N}$. Construct a $2mn^{k+1}\times2$ array, \textbf{S}, over $2mn^k$ roots of 
unity, where $\textbf{S} = [S_{i,j}] = \omega^{\left\lfloor i (i + j)/n\right\rfloor}$ and $\omega = e^{2 \pi \sqrt{-1}/(2mn^k)}$. 
The sequence \textbf{s} is constructed by enumerating, row-by-row, the array \textbf{S}.\\

We now show that \textbf{s} has perfect periodic autocorrelation. We show \textbf{s} is perfect
by showing that it has the array orthogonality property (AOP) for the divisor 2. First, we 
show that the cross-correlation of the two columns of \textbf{S} is zero for every non-zero shift. 
\begin{equation}
\theta_{S_{i,0}, S_{i,1}}(\kappa) = \sum_{i=0}^{2mn^{k+1}-1} S_{i,0}\, S_{i+\kappa, 1}^* \tag*{(1)}
\end{equation}

Let $i = q n + r$, ($r < n$), and $\kappa = q' n+r'$, ($r'<n$), then (1) becomes

\begin{align*}
\theta_{S_{qn+r,0}, S_{qn+r,1}}(q'n+r') &= \sum_{q=0}^{2mn^k-1}\sum_{r=0}^{n-1} S_{qn+r,0} S_{(q+q')n+r+r',1}^*\\
&= \sum_{q=0}^{2mn^k-1}\sum_{r=0}^{n-1} \omega^{\left\lfloor \frac{(qn+r)^2}{n}\right\rfloor}
	\omega^{-\left\lfloor \frac{\left((q+q')n + r+r'\right)^2 + (q+q')n+r+r'}{n}\right\rfloor}\\
&= \sum_{q=0}^{2mn^k-1}\sum_{r=0}^{n-1} \omega^{-(2nq'+2r'+1)q - 2q'r + 
	\left\lfloor\frac{r^2}{n}\right\rfloor - \left\lfloor\frac{(r+r')(r+r'+1)}{n}\right\rfloor} \\
&= \omega^{-2q'r'-q'}
	\left(\sum_{q=0}^{2mn^k-1}\omega^{-(2nq'+2r'+1)q}\right)\left(\sum_{r=0}^{n-1}
	 \omega^{ -2q'r + \left\lfloor\frac{r^2}{n}\right\rfloor - \left\lfloor\frac{(r+r')(r+r'+1)}{n}\right\rfloor} \right).
\end{align*}
The leftmost summation above is zero, as $-2nq'-2r'-1 \neq 0 \mod 2mn^k$ (since $-2nq'-2r'-1$ is odd
for all $n,q',r'$, whereas $2mn^k$ is even for all $n,q',r'$). Thus \textbf{s} satisfies the 
first condition of the AOP. \\

Now we show that \textbf{s} satisfies the second condition of the AOP. That is, for all non-zero shifts, we 
show that the sum of the periodic autocorrelations of both columns of \textbf{S} sums to zero. 
\begin{align*}
\theta_{S_{i,0}}(\kappa) + \theta_{S_{i,1}}(\kappa) &= \sum_{i=0}^{2mn^{k+1}-1} S_{i,0}\, S_{i+\kappa,0}^* + 
	\sum_{i=0}^{2mn^{k+1}-1} S_{i,1}\, S_{i+\kappa,1}^*\\
&= \sum_{i=0}^{2mn^{k+1}-1} \omega^{\left\lfloor \frac{i^2}{n} \right\rfloor} 
		\omega^{-\left\lfloor \frac{(i+\kappa)^2}{n} \right\rfloor} 
		+ \sum_{i=0}^{2mn^{k+1}-1} \omega^{\left\lfloor \frac{i^2+i}{n} \right\rfloor} 
		\omega^{-\left\lfloor \frac{(i+\kappa)^2 + i + \kappa}{n} \right\rfloor} \tag*{(2) + (3)}
\end{align*}

Let $i = q n + r$, ($r < n$), and $\kappa = q' n+r'$, ($r'<n$), then (2) becomes
\begin{align*}
&\omega^{-2q'r'-nq'^2} 
	\sum_{q=0}^{2mn^k-1}\sum_{r=0}^{n-1}
	\omega^{ -2(nq'+r')q - 2q'r + \left\lfloor\frac{r^2}{n}\right\rfloor - 
		\left\lfloor\frac{r^2+r'^2+2r'r}{n}\right\rfloor}\\
&= \omega^{-2q'r'-nq'^2}  
	\left(\sum_{q=0}^{2mn^k-1} \omega^{-2(nq'+r')q} \right)
	\left(\sum_{r=0}^{n-1} \omega^{- 2q'r + \left\lfloor\frac{r^2}{n}\right\rfloor - 
		\left\lfloor\frac{r^2+r'^2+2r'r}{n}\right\rfloor}\right). \tag*{(4)}
\end{align*}

Similarly, (3) becomes  
\begin{align*}
& \omega^{-2q'r' - n q'^2 - q'}
	\sum_{q=0}^{2mn^k-1}\sum_{r=0}^{n-1} 
		 \omega^{-2(nq'+r')q - 2q'r + \left\lfloor \frac{r^2+r}{n}\right\rfloor - 
		 	\left\lfloor \frac{r^2+2r'r+r+r'^2+r'}{n}\right\rfloor}\\
&= \omega^{-2q'r' - n q'^2 - q'}
	\left(\sum_{q=0}^{2mn^k-1} \omega^{-2(nq'+r')q} \right)
	\left(\sum_{r=0}^{n-1} \omega^{- 2q'r + \left\lfloor \frac{r^2+r}{n}\right\rfloor - 
		 	\left\lfloor \frac{r^2+2r'r+r+r'^2+r'}{n}\right\rfloor} \right). \tag*{(5)}
\end{align*}
Then $\theta_{S_{i,0}}(\kappa) + \theta_{S_{i,1}}(\kappa) = (4) + (5)$ is given by  \\
\leftline{$\omega^{-2q'r'-nq'^2} \left(\displaystyle\sum_{q=0}^{2mn^k-1} \omega^{-2(nq'+r')q} \right)
	\left(\displaystyle\sum_{r=0}^{n-1} \omega^{-2q'r + \left\lfloor\frac{r^2}{n}\right\rfloor - 
		\left\lfloor\frac{r^2+r'^2+2r'r}{n}\right\rfloor} +\right.$}
\rightline{$\left. \omega^{-q'}
	\displaystyle\sum_{r=0}^{n-1} \omega^{-2q'r + \left\lfloor \frac{r^2+r}{n}\right\rfloor - 
		 	\left\lfloor \frac{r^2+2r'r+r+r'^2+r'}{n}\right\rfloor} \right).$}
The summation $\sum_{q=0}^{2mn^k-1} \omega^{-2(nq'+r')q}$ is non-zero when 
$-2(nq'+r') = 0 \mod 2mn^k$, which is when $q'=-m n^{k-1}$, $r'=0$ (excluding $q'=r'=0$ as we 
only consider off-peak autocorrelations). In which case we have \\
\leftline{$\displaystyle\sum_{r=0}^{n-1} \omega^{-2q'r + \left\lfloor\frac{r^2}{n}\right\rfloor - 
		\left\lfloor\frac{r^2+r'^2+2r'r}{n}\right\rfloor}  = 
\displaystyle\sum_{r=0}^{n-1} \omega^{-2q'r + \left\lfloor \frac{r^2+r}{n}\right\rfloor - 
		 	\left\lfloor \frac{r^2+2r'r+r+r'^2+r'}{n}\right\rfloor} = $}
\rightline{$\displaystyle\sum_{r=0}^{n-1} \omega^{-2q'r} = \displaystyle\sum_{r=0}^{n-1} e^{\left(\frac{2 \pi \sqrt{-1}}{n}\right)r} = 0.$}
Thus, $\theta_{S_{i,0}}(\kappa) + \theta_{S_{i,1}}(\kappa) = 0$, so \textbf{s} satisfies the 
second condition of the AOP. It follows that \textbf{s} is a perfect sequence.\\

We note that the array, \textbf{S}, also has perfect periodic autocorrelation. The proof follows from the sequence, \textbf{s}, 
having the AOP. \\

In terms of the ratio of the sequence length to the number of phases, this construction sits below the construction of 
Milewski and above the constructions of Chu and Liu. 

\bibliographystyle{abbrv}

\begin{thebibliography}{99} 

\bibitem[Alltop, 1980]{Alltop1980} W. O. Alltop, ``Complex sequences with low periodic correlations", 
\textit{IEEE Trans. Inform. Theory}, vol. IT-26, no. 3, pp. 350-354, May 1980

\bibitem[Blake, 2012]{Blake2012} S. T. Blake, A. Z. Tirkel, ``A Construction for Periodic ZCZ Sequences", submitted 
for publication, December 2012

\bibitem[Chu, 1972]{Chu1972} D. C. Chu, ``Polyphase Codes With Good Periodic Correlation Properties", 
	\textit{IEEE transactions on information theory}, vol. 18, no. 4, pp. 531-532, July 1972

\bibitem[Farnett, 1990]{Farnett1990} E. C. Farnett et al, ``Pulse Compression Radar" \textit{Radar Handbook, 
2nd edition, Skolnik, M., Ed., McGraw-Hill}, 1990

\bibitem[Frank, 1962]{Frank1962}R. L. Frank, S. A. Zadoff and R. Heimiller, ``Phase Shift Pulse Codes with Good Periodic Correlation Properties", \textit{IRE Transactions on Information Theory}, vol. 8, no. 6, pp. 381-382, October 1961

\bibitem[Gabidulin, 1993]{Gabidulin1993} E. M. Gabidulin, ``Non-binary sequences with perfect periodic auto-correlation
and with optimal periodic cross-correlation", \textit{Proc. IEEE Int. Symp. Inform. Theory}, San Antonio, USA, pp. 412,
January 1993

\bibitem[Heimiller, 1961]{Heimiller1961} R. C. Heimiller, ``Phase Shift Pulse Codes with Good Periodic 
Correlation Properties", \textit{IRE Transactions on Information Theory}, vol. 7, no. 4, pp. 254-257, October 1961

\bibitem[Ipatov, 1979]{Ipatov1979} V. P. Ipatov, ``Ternary sequences with ideal autocorrelation properties",
\textit{Radio Eng. Electron. Phys.}, vol. 24, pp. 75-79, October 1979

\bibitem[Ipatov, 2005]{Ipatov2005} V. P. Ipatov, ``Spread Spectrum and CDMA: Principles and Applications",
\textit{John Wiley \& Sons}, 2005

\bibitem[Kumar, 1985]{Kumar1985} P. V. Kumar, R. A. Scholtz, L. R. Welch, ``Generalized Bent functions and 
their properties", \textit{J. Combinat. Theory, series A}, vol. 40, no. 1, pp. 90-107, 1985

\bibitem[Lewis, 1982]{Lewis1982} B. L. Lewis, F. F. Kretschmer, ``Linear frequency modulation derived polyphase
pulse compression", \textit{IEEE Trans. on AES}, vol. AES-18, no. 5, pp. 637-641, September 1982

\bibitem[Liu, 2004]{Liu2004} Y. Liu and P. Fan, ``Modified Chu sequences with smaller alphabet size", \textit{Electronics Letters}, vol. 40, no. 10, May 2004

\bibitem[Milewski, 1983]{Milewski1983} A. Milewski, ``Periodic Sequences with Optimal Properties for Channel 
Estimation and Fast Start-Up Equalization", \textit{IBM Journal of Research and Development}, vol. 27, no. 5, 
pp. 426-431, September 1983

\bibitem[Mow, 1993]{Mow1993} W. H. Mow, ``A Study of Correlation of Sequences", PhD, Department of Information 
Engineering, The Chinese University of Hong Kong, 1993

\bibitem[Simon, 1985]{Simon1985} M. K. Simon, ``Spread spectrum communications, Volume 1", 
\textit{Computer Science Press, The University of Michigan}, 1985

\bibitem[Tirkel, 1993]{Tirkel1993} A. Z. Tirkel, G. A. Rankin, R. M. Van Schyndel, W. J. Ho, N. R. A. Mee, 
	C. F. Osborne,  ``Electronic Water Mark", DICTA 93, Macquarie University, pp. 666-673, 1993

\bibitem[Van Schyndel, 2001]{VanSchyndel2001} R. G. Van Schyndel ``Using Phase-Modulated Probe Signals to 
Recover Delays from Higher Order Non-linear Systems", \textit{IEEE Engineering in Medicine and Biology: 
Biomedical Research in 2001}, pp. 94-97, February 2001

\bibitem[Xu, 2011]{Xu2011} L. Xu, ``Phase coded waveform design for Sonar Sensor Network", 
\textit{Conference on Communications and Networking in China (CHINACOM), 2011 6th International ICST},
pp. 251-256, August 2011

\end{thebibliography}

\end{document}